\documentclass{iaus}
\usepackage{graphicx}

\title[IAU 297.~~First results of a new DIB survey] 
{First results from a study of DIBs with thousands of high-quality massive-star spectra}

\author[J. Ma\'{\i}z Apell\'aniz \etal]   
{J. Ma\'{\i}z Apell\'aniz$^{1,*}$,
A. Sota$^1$,
R. H. Barb\'a$^2$,
N. I. Morrell$^3$, \\
A. Pellerin$^4$,
E. J. Alfaro$^1$,
\and
S. Sim\'on-D{\'\i}az$^5$}

\affiliation{$^1$Instituto de Astrof{\'\i}sica de Andaluc{\'\i}a-CSIC, Spain; $^2$Universidad de La Serena, Chile; \\
$^3$Las Campanas Observatory, Chile; $^4$Mount Allison University, Canada; \\
$^5$Instituto de Astrof{\'\i}sica de Canarias, Spain; $^*$e-mail: {\tt jmaiz@iaa.es}}

\pubyear{2013}
\volume{297}  
\pagerange{1--3}
\jname{The Diffuse Interstellar Bands}
\editors{J. Cami \& N. Cox, eds.}
\begin{document}

\maketitle

\begin{abstract}
We are using five different surveys to compile the largest sample of diffuse interstellar band (DIB)
measurements ever collected. GOSSS is obtaining intermediate-resolution blue-violet spectroscopy of $\sim$2500 
OB stars, of which 60\% have already been observed and processed. The other four surveys have already collected
multi-epoch high-resolution optical spectroscopy of 700 OB stars with different telescopes, including the 9 m 
Hobby-Eberly Telescope in McDonald Observatory. Some of our stars are highly-extinguished targets for which no 
good-quality optical spectra have ever been published. For all of the targets in our sample we have obtained 
accurate spectral types, measured non-DIB ISM lines, and compiled information from the literature to calculate 
the extinction. Here we present the first results of the project, the properties of twenty DIBs in the 
4100-5500 \AA\ range. We clearly detect a couple of previously elusive DIBs at 4170~\AA\ and 4591~\AA; the latter
could have coronene and ovalene cations as carriers.
\end{abstract}

\firstsection 
\section{The project}

The Galactic O-Star Spectroscopic Survey (GOSSS, Ma{\'\i}z Apell\'aniz \etal\ 2011, Sota \etal\ 2011) is obtaining
long-slit, $R\sim$2500, blue-green spectroscopy of $\sim$2500 massive stars in both hemispheres, including all known
O stars with $B < 12$. Its main purpose is to characterize the O-star population in the solar neighborhood by providing accurate
spectral types for all of the observed targets. We currently have spectra for 1592 stars and we plan to reach 2500 within two years.

Four other surveys are obtaining high-resolution spectroscopy of a subsample of the GOSSS stars with the purposes of doing
detailed atmospheric modeling and calculating the orbits of the spectroscopic binaries. Three of those surveys,
OWN (Barb\'a \etal\ 2010), IACOB (Sim\'on-D{\'\i}az \etal\ 2011a, 2011b), and NoMaDS (Ma{\'\i}z Apell\'aniz \etal\ 2012), have been
described elsewhere. The fourth one, CAF\'E-BEANS (Calar Alto Fiber-fed \'Echelle Binary Evolution Andalusian Survey, P.I.:
Ignacio Negueruela), is obtaining multi-epoch $R=$ 65\,000 spectroscopy of northern stars using the CAF\'E spectrograph at the 
2.2~m telescope at Calar Alto (Aceituno \etal\ 2012).

The original goal of these surveys was to study the stars but they also contain an unprecedented amount of information on 
optical ISM lines. That led us to start a parallel project to obtain and process such information. In this first analysis we study
the properties of the DIBs seen in the GOSSS data.


\hspace{-1cm}\begin{minipage}[t]{0.59\textwidth}
{\bf Table 1.} GOSSS extincted and reference stars used in this work.
\vspace*{2mm}

\begin{tabular}{cllr@{}l}
ID                                 &
\multicolumn{1}{c}{Extincted star} &
\multicolumn{1}{c}{Spectral type} 	&
\multicolumn{2}{c}{$E(B-J)$}       \\
\hline
1 & Cyg OB2-12	             & B5 Ia	      & $\;\;\;\;$10&.38	\\
2 & ALS 19\,626	            & B0 Ia	      &            8&.61	\\
3 & 2MASS                  	& B0 Ia	      &            8&.33	\\
  & $\,$ J20333821+4041064	 &             &                 	\\
4 & ALS 21\,079	            & O7 Ib-II(f)	&            7&.64	\\
5 & Cyg OB2-22 B	           & O6 V((f))	  &            7&.56	\\
6 & Cyg OB2-22 A	           & O3 If*	     &            7&.51	\\
7 & ALS 15\,114	            & O7.5 Vz	    &            7&.50	\\
8 & 2MASS                   & O6.5 III(f)	&            7&.12	\\
  & $\;$ J20344410+4051584  &            	&                 	\\
9 & ALS 18\,747	            & O5.5 Ifc	   &            5&.94	\\
\hline
ID                                 &
\multicolumn{1}{c}{Reference star} &
\multicolumn{1}{c}{Spectral type} 	&
\multicolumn{2}{c}{$E(B-J)$}       \\
\hline
1 & $\eta$ CMa	             & B5 Ia	      &            0&.37 \\
2 & HD 91\,969	             & B0 Ia	      &            0&.82 \\
3 & HD 91\,969	             & B0 Ia	      &            0&.82 \\
4 & HD 94\,963	             & O7 II	      &            0&.68 \\
5 & HDE 303\,311	           & O6 V	       &            1&.53 \\
6 & $\zeta$ Pup	            & O4 I(n)f	   &            0&.07 \\
7 & HD 53\,975	             & O7.5 Vz	    &            0&.62 \\
8 & HD 152\,723 AaAb	       & O6.5 III(f)	&            1&.45 \\
9 & HD 93\,632	             & O5.5 Ifc	   &            2&.33 \\
\hline
\end{tabular}
\end{minipage} $\;$
\begin{minipage}[t]{0.54\textwidth}
{\bf Table 2.} Properties of the DIBs measured in this work using Gaussian fits. The first column gives the group ID (some DIBs were 
fitted together), the next two give the fit results, the fourth one the average equivalent widths of the used stars, and the last one the 
number of stars used.
\vspace*{1mm}

\begin{tabular}{rcr@{}lcc}
 gr. & $\lambda_{\rm c}$ (\AA) & \multicolumn{2}{c}{FWHM (\AA)} & EW (\AA)      & $n_{\rm st}$ \\
\hline
 1   & 4179.48$\pm$0.61        & 21.88&$\pm$1.73                & 0.62$\pm$0.14 & 8            \\
 2   & 4427.94$\pm$0.11        & 24.15&$\pm$0.30                & 3.55$\pm$0.68 & 9            \\
 3   & 4501.67$\pm$0.09        &  3.24&$\pm$0.40                & 0.17$\pm$0.04 & 9            \\
 4   & 4591.19$\pm$0.56        & 24.95&$\pm$1.58                & 0.68$\pm$0.25 & 6            \\
 5   & 4726.70$\pm$0.05        &  3.88&$\pm$0.30                & 0.28$\pm$0.05 & 9            \\
 6   & 4761.12$\pm$0.27        & 19.72&$\pm$1.06                & 0.58$\pm$0.09 & 9            \\
 6   & 4762.36$\pm$0.09        &  2.69&$\pm$0.48                & 0.10$\pm$0.01 & 9            \\
 6   & 4779.69$\pm$0.21        &  5.48&$\pm$0.57                & 0.13$\pm$0.02 & 9            \\
 7   & 4879.83$\pm$0.16        & 11.51&$\pm$0.63                & 0.43$\pm$0.07 & 7            \\
 7   & 4887.43$\pm$0.65        & 39.75&$\pm$1.14                & 2.16$\pm$0.36 & 7            \\
 8   & 4963.85$\pm$0.15        &  2.62&$\pm$0.66                & 0.06$\pm$0.02 & 8            \\
 9   & 4984.59$\pm$0.23        &  1.33&$\pm$1.29                & 0.02$\pm$0.01 & 8            \\
10   & 5109.66$\pm$0.43        & 14.91&$\pm$1.34                & 0.23$\pm$0.08 & 6            \\
11   & 5155.99$\pm$0.49        & 15.97&$\pm$1.50                & 0.14$\pm$0.03 & 7            \\
12   & 5236.29$\pm$0.24        &  2.27&$\pm$0.77                & 0.06$\pm$0.02 & 5            \\
12   & 5245.43$\pm$0.57        &  7.26&$\pm$1.56                & 0.08$\pm$0.03 & 5            \\
13   & 5363.52$\pm$0.13        &  2.74&$\pm$0.55                & 0.07$\pm$0.01 & 7            \\
14   & 5449.83$\pm$0.22        & 14.06&$\pm$0.61                & 0.48$\pm$0.13 & 8            \\
15   & 5487.23$\pm$0.12        &  6.63&$\pm$0.33                & 0.35$\pm$0.05 & 6            \\
15   & 5494.29$\pm$0.26        &  1.90&$\pm$0.96                & 0.04$\pm$0.01 & 6            \\
\hline
\end{tabular}
\end{minipage}

\vspace*{-1.2cm}

\section{4100-5500 \AA\ DIB properties}

We selected 9 stars observed by GOSSS with high extinction, good S/N spectra, and coverage of the full 3900-5500 \AA\ range
(Table 1). For each star we selected a corresponding GOSSS star with low extinction and similar spectra type in order to apply the
traditional pair method, in which the low-extinction spectrum is subtracted from the high-extinction one in order to eliminate the
stellar contribution. The spectra were put in the ISM reference system using the Ca\,{\sc ii}~$\lambda$3934 line. The average
profile was then calculated for each DIB by selecting the stars in the sample with the largest EWs, normalizing by it, and
calculating the mean and standard deviation of the spectra at each wavelength. The result was fit using either Gaussian and
Lorentzian profiles (multiple in the case where DIBs overlap so they must be fitted in groups). Finally, the instrumental width was
subtracted to the fit FWHM. Results for the Gaussian fits are shown in Table 2.

\begin{itemize}
 \item We clearly detect the seldom seen 4179 \AA\ broad DIB (Knoechel \& Moffat 1982, Jenniskens \& D\'esert 1994; Fig. 1).
 \item The broad intense 4428 \AA\ DIB is better fit by a Lorentzian than a Gaussian (see Snow \etal. 2002). However, the 
       differences are small and the broad Lorentzian wings make rectification difficult in practice, so Gaussian fits are 
       less noisy for most data.
 \item We detect the previously elusive 4591 \AA\ broad DIB, which could be produced by coronene 
       (C$_{24}$H$_{12}$) and ovalene (C$_{32}$H$_{14}$) cations (Ehrenfreund et al. 1995, Fig. 1).
 \item The 4770 \AA\ region requires a fit with two narrow and one broad DIBs (Fig. 1).
 \item The 4880 \AA\ region requires a fit with one broad and one intermediate DIBs (Fig. 1). 
 \item The 5110 \AA\ DIB shows a possible additional component in its left wing. 
 \item There may be a broad component around the 5364 \AA\ DIB (not measured).
 \item At this resolution the two narrow DIBs to the left and right of He\,{\sc ii}~$\lambda$5412 are blended with the stellar line 
       and are not analyzed here.
 \item The 5240 \AA\ region shows two clearly separated components at this resolution.
 \item The 5490 \AA\ region shows two clearly separated components at this resolution.
\end{itemize}

\begin{figure}[b]
\vspace*{-3mm}
\begin{center}
 \includegraphics[width=0.47\linewidth]{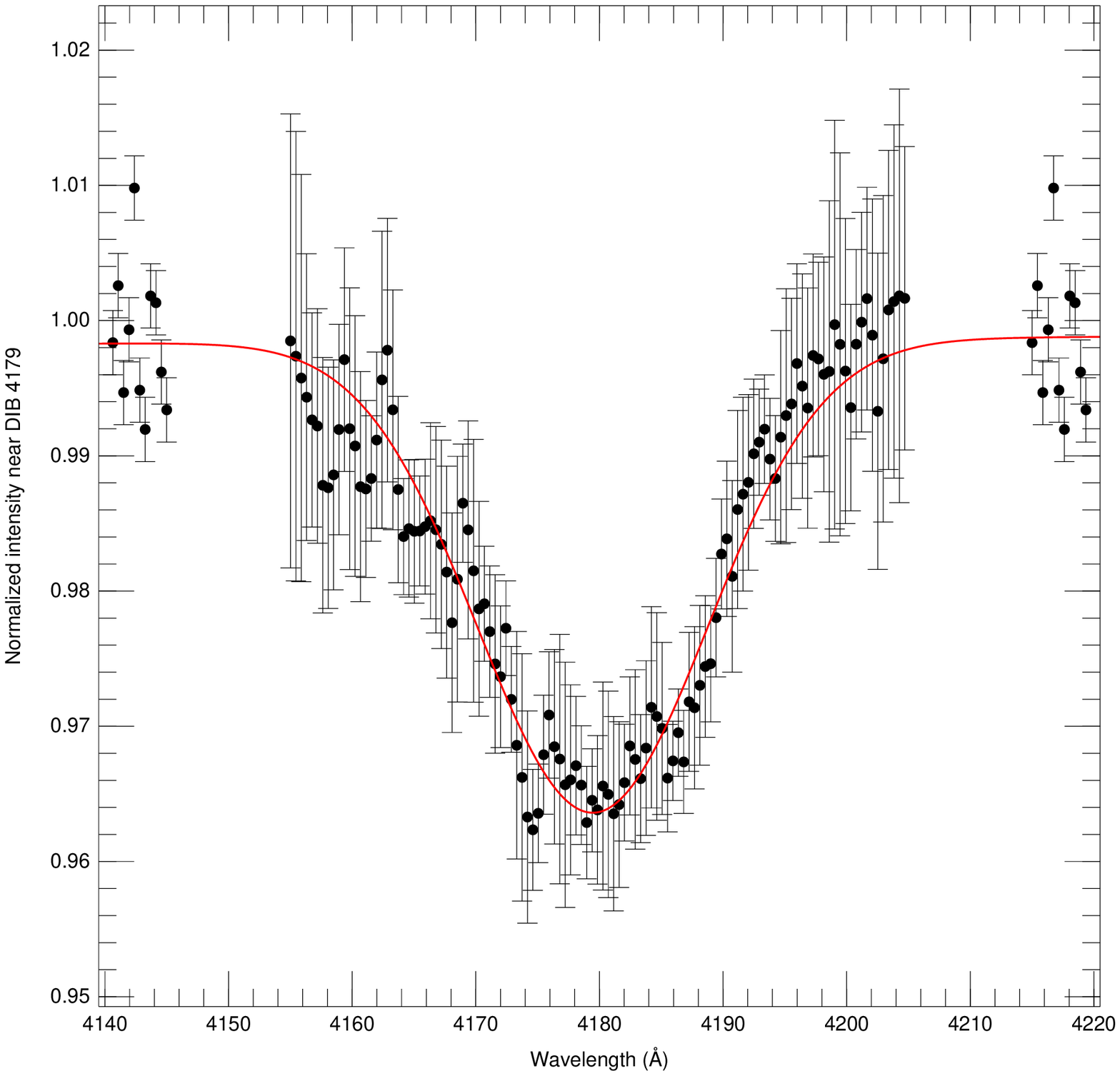} \
 \includegraphics[width=0.47\linewidth]{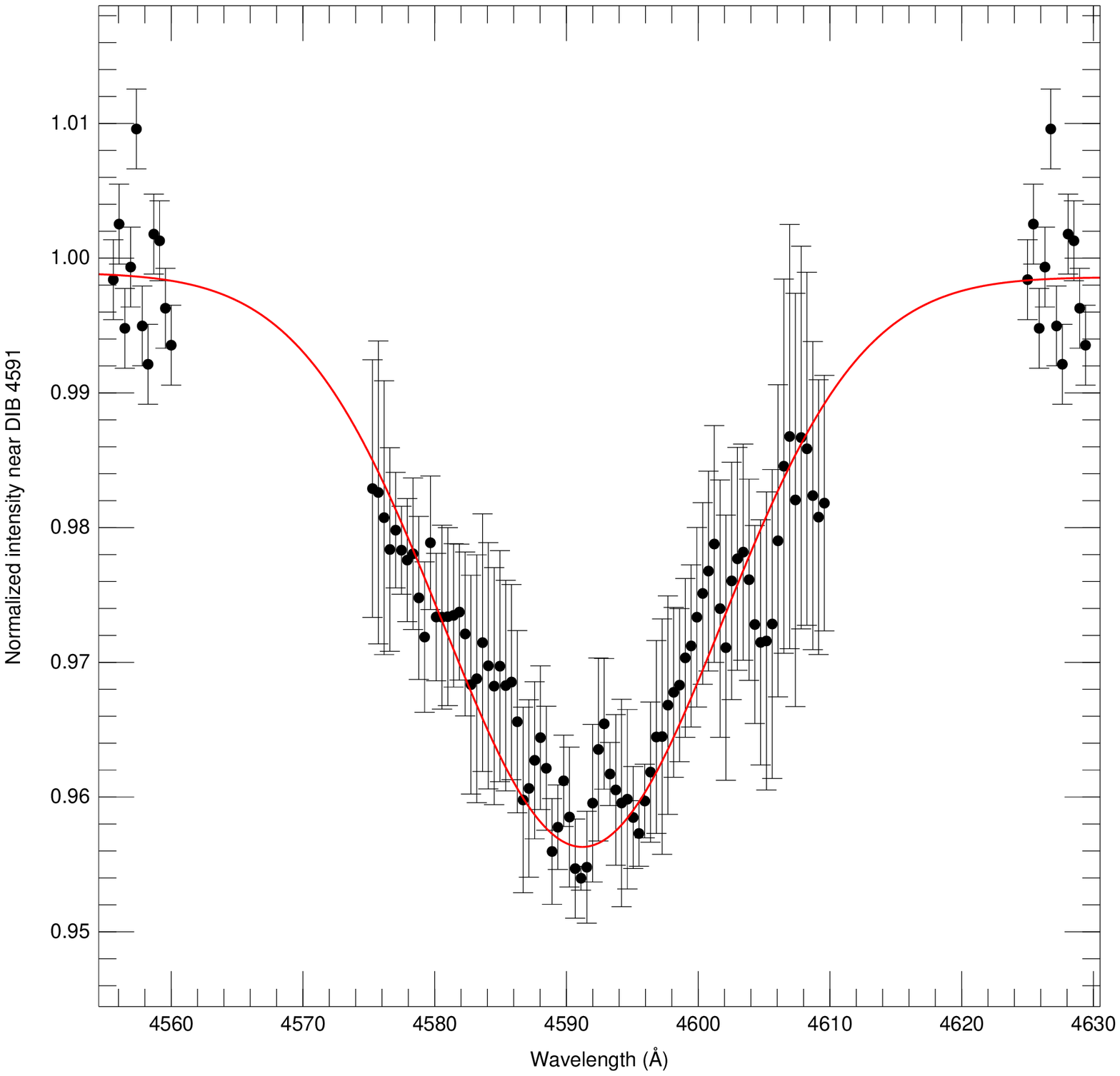}

\vspace*{-2mm}
 \includegraphics[width=0.47\linewidth]{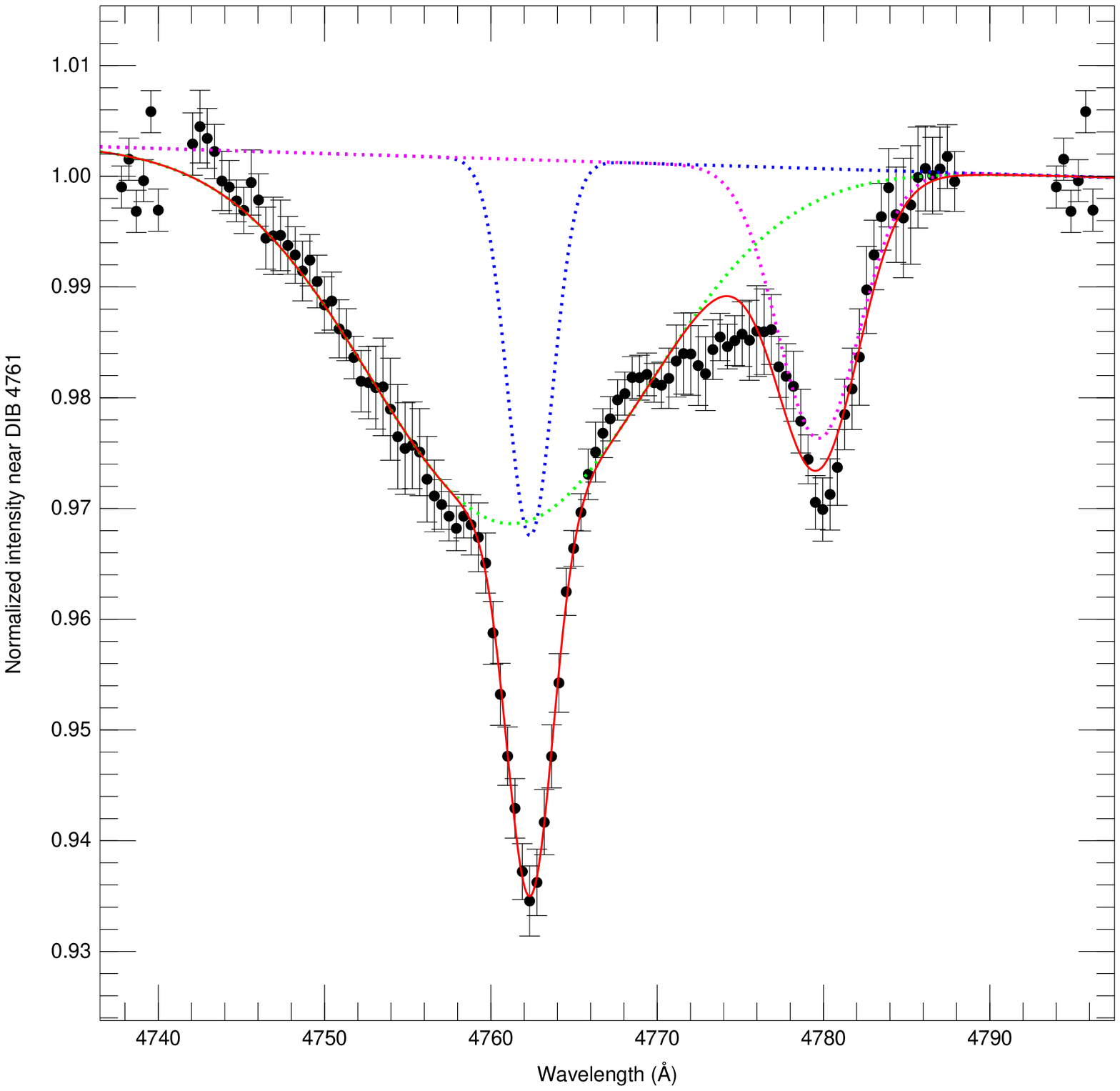} \
 \includegraphics[width=0.47\linewidth]{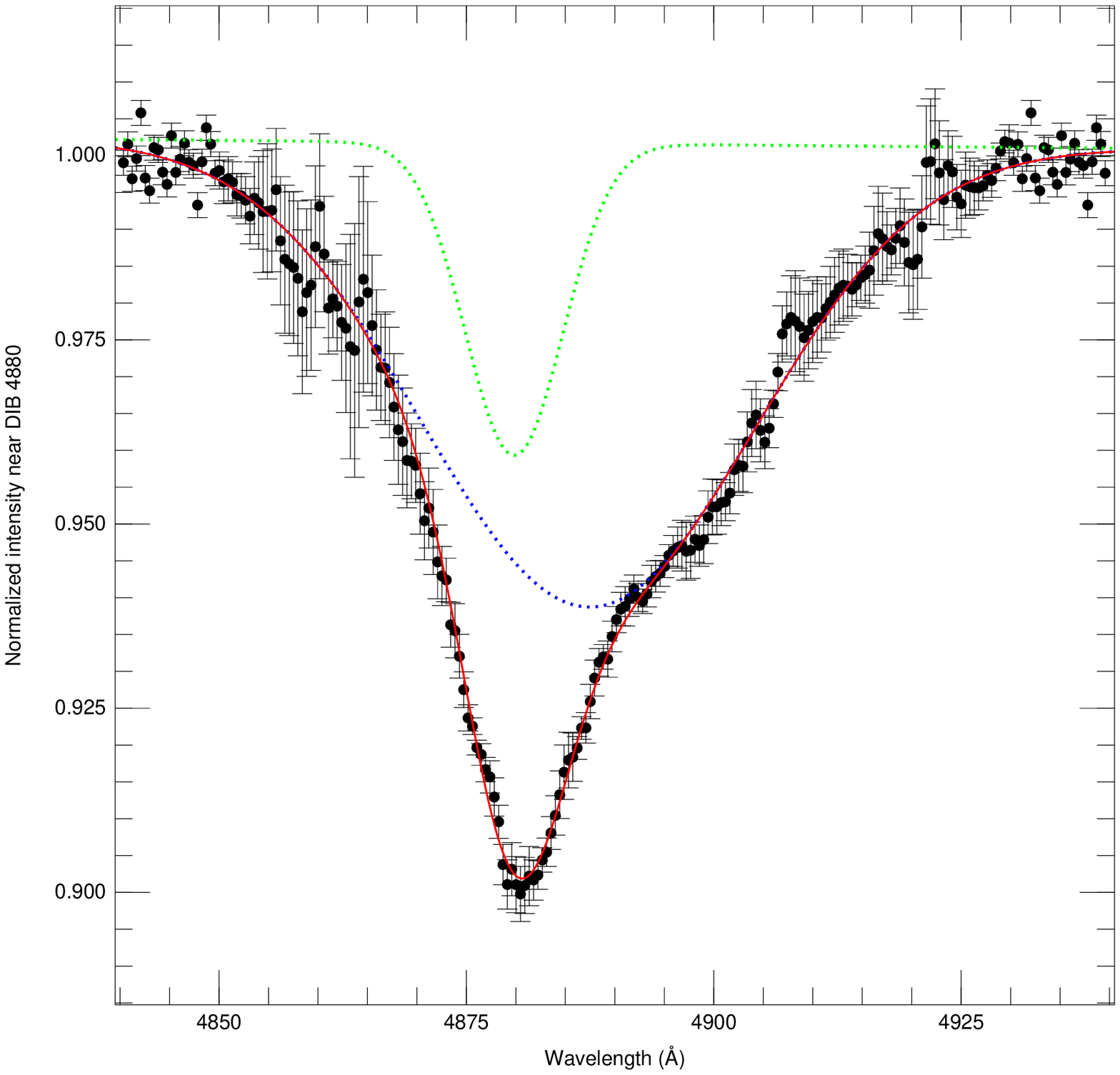}
\vspace*{-3mm}
 \caption{Gaussian fits for DIB groups 1, 4, 6, and 7. Error bars show the standard deviation of the average profile normalized
          by EW. Dotted lines are used for individual components.}
   \label{fig1}
\end{center}
\end{figure}

\end{document}